# Three-dimensional modeling magneto-optical trapping of MgF molecule with multilevel rate equations

Supeng Xu,[1] Meng Xia,[1] Ruoxi Gu,[1] Yanning Yin,[1] Liang Xu,[1] Yong Xia,[1,2*] and Jianping Yin[1]

[1] State key laboratory of precision spectroscopy, School of Physics and Materials Science, East China Normal University, Shanghai 200062, China

[2] NYU-ECNU Institute of Physics at NYU Shanghai, Shanghai 200062, China

*yxia@phy.ecnu.edu.cn

**Abstract**

We present a theoretical study of magneto-optical trapping (MOT) force exerted on magnesium monofluoride (MgF) with 3D rate equations, in which we have considered the complex vibrational and rotational levels and the effects of small internal splittings and degeneracies, including fine and hyperfine structures and the magnetic quantum numbers. We investigate the feasibility of MOT for MgF with a very small excited state $g$-factor ($g_e = -0.0002$) and a large radiative decay rate ($\Gamma = 2\pi \times 22$MHz) for the electronic transition of $X^2\Sigma^+$ to $A^2\Pi_{1/2}$ states. We also optimize the MOT with reference to the three-, four- and more-frequency component models with various polarization configurations and detunings. By applying the dual frequency arrangement to more than one hyperfine level, we suggest a configuration of the 3+1 frequency components for achieving the MOT of MgF.

## I. Introduction

Development of a molecular MOT should really mirror the huge historical success achieved by the atomic MOT [1]. Realizing such a powerful technique for producing a diverse set of dense, ultracold diatomic molecular species opens a new chapter for molecular science and it will greatly advance understandings in precision measurement, complex quantum systems under precise control and ultracold chemistry in the most fundamental way [2-4]. The dawn of ultracold polar molecules was signaled with the production of ground-state polar molecules KRb near quantum degeneracy in 2008 [5]. In that experiment they bypassed the problem of direct cooling of molecules, taking advantage of ultracold atoms and then using resonant association techniques for producing ground-state molecules. This approach is currently limited to bi-alkalies, such several recently created ultracold polar molecules also include RbCs [6,7], NaK [8], and NaRb [9]. On the other hand, tremendous progress has been made in direct laser cooling and the MOT of diatomic molecules, i.e. SrF [10,11], YO [12], CaF [13,14] and triatomic molecule SrOH [15], even polyatomic molecule $CH_3F$ [16], $H_2CO$ [17]. In addition, some other ongoing candidates, such as YbF [18], BaF [19, 20], BaH [21], have attracted great interest as well. To date, the temperature of the cooled diatomic molecule is well below the Doppler limit [13,22,23]. The maximum number of molecules, $1.0 \times 10^5$, were captured by L. Anderegg *et al.* through radio frequency CaF MOT [14].

Magnisum monofluoride, due to its highly diagonal Franck-Condon factors and strong spontaneous radiation decay, can also be a good candidate for molecular MOT [24]. In general, the magneto-optical trapping force is very weak if the excited-state *g*-factor of the laser-cooling molecule, $g_e$, is much smaller than the ground-state *g*-factor, $g_g$. Fortunately, due to "the dual-frequency effect" (the sublevel involved in the transition is addressed by two frequency components with different polarization at the same time, which is an effective way to eliminate the effect of the dark states) caused by the multiple levels in the ground state of the molecule, molecular force is no longer negligible compared to the atomic one [25]. However, note that MgF has a much smaller excited state $g_e$ ($g_e \approx -0.0002$), compared with CaF ($g_e \approx -0.021$) and SrF ($g_e \approx -0.088$) [26,27]. Also, the unique hyperfine structure of the ground state [24] (the hyperfine level interval between the upper $F = 2$ and $F = 1$ levels of the ground state is ~0.4 $\Gamma$, which may break down the dual-frequency). So, it's necessary for us to verify whether MgF is appropriate to MOT or not and select the optimal polarization configuration.

In this paper, we apply three dimension (3D) multilevel rate equations with multiple frequencies of laser to model the MOT of MgF molecule for the $A^2\Pi_{1/2} - X^2\Sigma^+$ transition. The dual-frequency mechanism is considered and we focus mainly on the choice of laser polarization and detuning. Throughout the discussion, we do not take the vibrational repump transitions into consideration, since the influence on the MOT is small. Our results show that three-frequency component can cool molecule to a lower temperature while the four-frequency component is preferred in trapping molecules. Moreover, by adding one extra frequency component, both the maximum damping force and the relatively large trapping force can be obtained at a cost of the capture velocity. These results will be an effective guide for our experiment.

## II. Modeling MOT of MgF molecule

### A. Rate equations

We apply the approach of multilevel rate equations which include "the dual-frequency" effect to model the MOT of MgF molecule [25,28]. The molecule has a set of ground states, g, and excited states, e, with populations $n_g$ and $n_e$ respectively, interacting with a laser field with components, p. Each laser component has an angular frequency $\omega_p$ and propagates in the direction of the unit vector $\kappa_p$. The frequencies are similar, so we use a single wavelength $\lambda \approx 2\pi c/\omega_p = 359.3$ nm for all components[29]. There is a quadrupole magnetic field represented by B = A ($x\hat{x}, y\hat{y}, -2z\hat{z}$), where $\hat{x}$, $\hat{y}$, $\hat{z}$ are unit vectors in the *x*, *y*, *z* axes, and A is the field gradient in the $xy$ plane. According to our modeling, the magnetic field gradient will mainly influence the position of the peak, so we set A = 30 G/cm and A = 10 G/cm for

three- and four-frequency configurations, respectively. This ensures the minimum value of the trapping force within the beam waist radius.

The intensity distribution of each laser beam is Gaussian:

$$I = \frac{2P}{\pi w^2} \exp\left(-\frac{2r^2}{w^2}\right) (r \leq r_t) \qquad (1)$$

Where $r$ is the distance from the center of the beam, $w$ is the $1/e^2$ radius, and P is the power of the beam. Taking into account the experimental feasibility, $w$ is set to 12 mm in the following discussion.

All excited states share one decay rate $\Gamma$. The polarization of the laser is resolved into components $(\sigma^-, \pi, \sigma^+)$ in the molecule's local coordinates, with relative amplitudes $(\frac{1}{2}, \frac{1}{\sqrt{2}}, \frac{1}{2})$, where $z$-axis is determined by the magnetic field direction. The molecules move slowly enough so they adiabatically follow the changes in the field direction. The rate equations for the system are:

$$\dot{r} = v, \qquad (2a)$$

$$\dot{v} = \frac{h}{m\lambda} \sum_{e,g,p} k_p R_{e,g,p}(n_g - n_e), \qquad (2b)$$

$$\dot{n}_g = \Gamma \sum_e f_{e,g} n_e - \sum_{e,p} R_{e,g,p}(n_g - n_e), \qquad (2c)$$

$$\dot{n}_e = -\Gamma n_e + \sum_{g,p} R_{e,g,p}(n_g - n_e), \qquad (2d)$$

$$\dot{\gamma} = \Gamma \sum_e n_e. \qquad (2f)$$

Here, $m$ is the mass of the molecule, $r$ and $v$ are the position and velocity of the molecule, respectively. $\Gamma$ is the decay rate, $\gamma$ is the number of scattered photons and $f_{e,g}$ is the branching ratio for spontaneous decay for the e-g transition, $R_{e,g,p}$ is the excitation rate between levels e and g driven by laser component p, which is,

$$R_{e,g,p} = \frac{\Gamma}{2} \frac{q_{e,g,p} s_p}{1 + 4(\delta_{e,g,p} - 2\pi k_p \cdot v/\lambda - \Delta\omega_{e,g})^2/\Gamma^2}, \qquad (3)$$

Where $s_p$ is the saturation parameter, $q_{e,g,p}$ is the fractional strength of the transition being driven, $\delta_{e,g,p} = \omega_p - \omega_{e,g}$ is the detuning from the resonance angular frequency for a stationary particle at zero field and $\Delta\omega_{e,g}$ is the Zeeman shift of the transition. For small magnetic fields, $\Delta\omega_{e,g} = (g_e M_e - g_g M_g)\mu_B B/\hbar$, where $g_e, g_g$ are the g-factors and $M_e, M_g$ are the magnetic quantum numbers of the excited and ground levels, respectively. The saturation parameter is $s_p = I_p/I_{sat}$, where $I_p$ is the intensity of laser component p, and $I_{sat} = \pi hc\Gamma/(3\lambda)^3$ is the saturation intensity for a two-level atom. The transition strength is $q_{e,g,p} = \frac{|\langle g|\hat{d}\cdot\epsilon_p|e\rangle|^2}{\Sigma_{g'}|\langle g'|\hat{d}|e\rangle|^2}$, where $\hat{d}$ is the dipole moment operator, and $\epsilon_p$ is the laser polarization. From these definitions, $q_{e,g,p} s_p = 2\Omega_{e,g}^2/\Gamma$, where $\Omega_{e,g}$ is the Rabi frequency at which the e-g transition is driven.

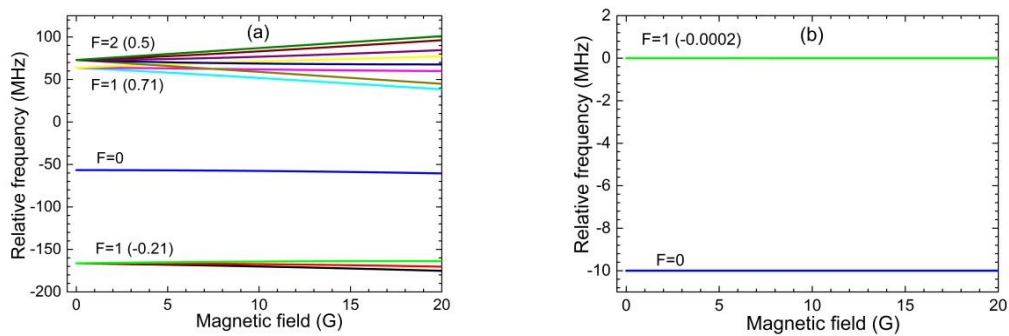

FIG. 1. Zeeman shifts of the states in magneto-optical trapping of MgF. (a) The four hyperfine components of the $X^2\Sigma^+$ ($v = 0$, $N = 1$) state. (b) The two hyperfine components of the $A^2\Pi_{1/2}$ ($v = 0$, $J = 1/2$) state with positive parity (e parity).

**B. Zeeman splitting of $X^2\Sigma^+$ and $A^2\Pi_{1/2}$ states**

To demonstrate the real MOT force, we take the full nonlinear Zeeman splitting of the ground states. Fig. 1(a) shows the relevant energy levels of the X state up to 20

G used in the equations. As seen, the X ($v = 0$, $N = 1$) state is split into four components due to the spin-rotation and hyperfine interactions, which are $F = 1, 0, 1$ and 2, respectively. The g-factors of the ground states, $g_g$, are given in parentheses after the F labels in Fig. 1(a). In the MOT modeling, all four components are addressed. For $A$ ($v = 0$, $J = 1/2$, +) state, we use linear Zeeman shifts given by their g-factors as showed in parentheses in Fig. 1(b). Note that the $F = 1$ level is made up of three Zeeman sublevels though it can't be distinguished because of the much smaller g-factor. The interval between the $F = 1$ and $F = 0$ levels was set to 10 MHz since the hyperfine splitting of A state is unknown and indistinguishable. Within the natural width $\Gamma$ ($2\pi \times 22$MHz) of $A^2\Pi_{1/2}$ state, the interval has minimal influence on our modeling results.

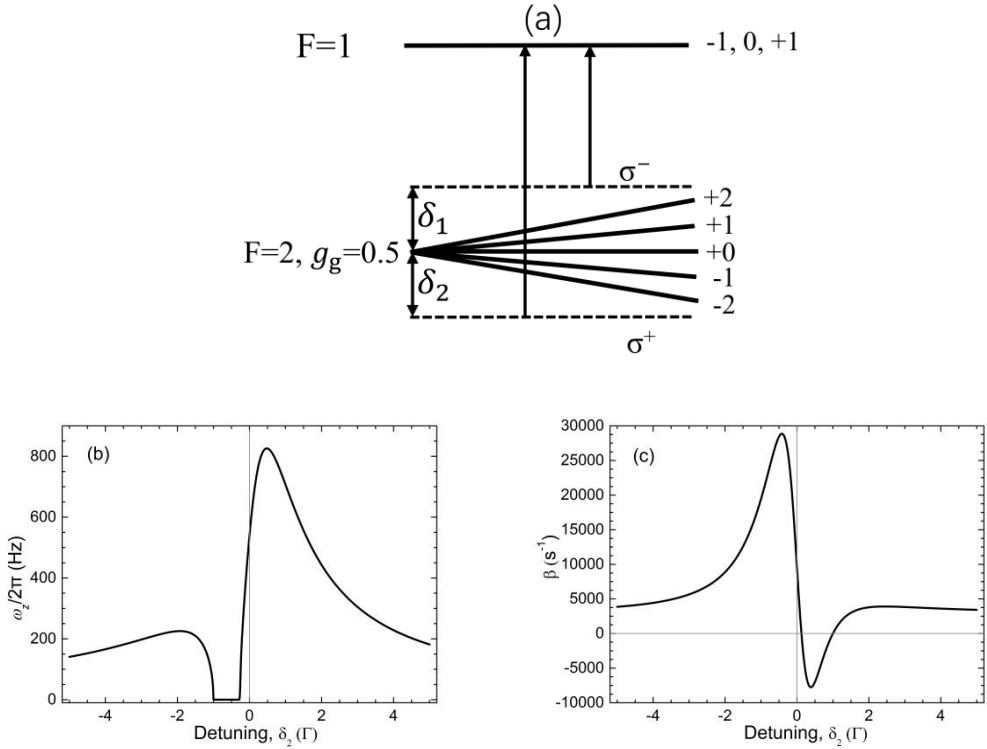

FIG. 2. (a) Illustration of the dual-frequency with ground level $F_g = 2, g_g = 0.5$ and excited level $F_e = 1, g_e = 0$. Two transitions with oppositely polarized frequency components were driven, and the detunings are $\delta_1$ and $\delta_2$, respectively. (b) Trap frequency versus $\delta_2$. (c) Damping coefficient versus $\delta_2$.

## C. The dual-frequency effect

We also consider the dual-frequency effect in the system in Fig. 2(a). The system consists of a ground level with $F_g = 2, g_g = 0.5$, an excited level with $F_e = 1, g_e = 0$ and two kinds of oppositely polarized frequency with detuning $\delta_1$ and $\delta_2$. The wavelength, mass and decay rate are all set equal to the MgF system. The molecule interacts with six orthogonal laser beams, each of which contains two frequencies. The power of each frequency is set to 40 mW, whose saturation parameter is ~0.28. $\delta_1$ value is fixed to $-\Gamma$, and $\delta_2$ value is varied. An effective way to demonstrate the trapping and cooling force of the MOT is to calculate the acceleration of a stationary molecule versus its displacement along the $z$ axis, and the acceleration of a molecule at the center of the MOT versus its speed in the $z$ direction. For small values of the displacement, $z$, and speed, $v_z$, we can write the acceleration as $a_z = -\omega_z^2 z - \beta v_z$, where $\omega_z/2\pi$ is the trap frequency and $\beta$ is the damping coefficient. Both of them can be used to characterize the MOT.

Figure 2(b) shows the trap frequency versus $\delta_2$. Here, we make a brief description of the results. The restoring force has a maximum at about $0.5\,\Gamma$ and remains significant for large positive detuning. There is no trapping when $\delta_2$ varies from $-\Gamma$ to $-0.3\,\Gamma$. However, there is still a considerable trapping force for $\delta_2 < -\Gamma$. These characteristics are attributed to the Zeeman splitting combined with multi-frequency lasers, which are named "the dual-frequency" [25].

Figure. 2(c) shows the dependence of the damping coefficient on $\delta_2$. Cooling effect occurs when $\beta$ is positive. When $\delta_2 < 0$, it is always cooling. While $0 < \delta_2 < \Gamma$, it is heating instead. When $\delta_2 > \Gamma$, it is cooling again.

## III. Cooling and trapping force in MOT

**TABLE I**: From left to right are the energy splitting of the hyperfine structure of X(0) and the frequency of each component in the three- and four-frequency schemes, respectively.

| F | X(0) (MHz) | Three-frequency scheme (MHz) | | Four-frequency scheme (MHz) | |
|---|---|---|---|---|---|
| 1 | -166.4 | f1 | -166.4 | f1 | -166.4 |
| 0 | -56.6 | f2 | -56.6 | f2 | -56.6 |
| 1 | 63.7 | f3 | 68.3 | f3 | 63.7 |
| 2 | 72.9 | | | f4 | 72.9 |

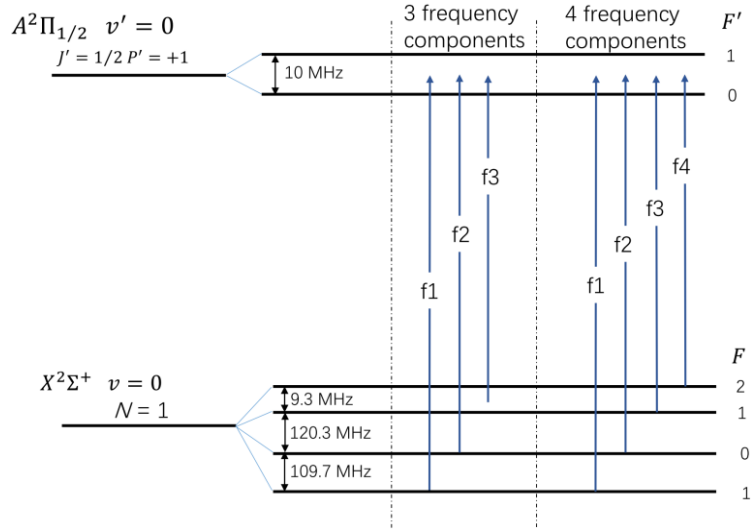

FIG. 3. Schemes for driving the *A*(0) - *X*(0) transition in a MOT of MgF. In the three-frequency configuration, f1, f2 and f3 are accordingly relative to the $F = 1$, $F = 0$ and the upper $F = 1, 2$ levels. For four-frequency case, f1, f2, f3 and f4 are relative to the $F = 1, 0, 1$ and 2 state, respectively.

Now, we move to the model of MOT for the *A*(0) - *X*(0) transition of MgF molecule. This transition has $\lambda = 359.3$ nm and $\Gamma = 2\pi \times 22$ MHz. The branching ratios of $A$ ($v = 0$, $J = 1/2$, +) - $X$ ($v = 0$, $N = 1$) transition were calculated by Yang [30]. Since the interval between the upper $F = 2$ and $F = 1$ is 9.3 MHz, which is less than Gamma ($\Gamma$), either three or four frequencies can be used to drive the four hyperfine components of the transition and the specific structures are shown in Fig. 3. From bottom to top, the intervals between the hyperfine level are, in $\Gamma$, 5.0, 5.5 and 0.4, which make MgF molecule a suitable molecule for the dual-frequency. Table I lists the relative frequencies of these components. In our model for the three-frequency configuration, f1 and f2 frequencies drive the $F = 1$ and $F = 0$ transition, respectively, and the upper $F = 1, 2$ transition is addressed by f2 and f3 frequencies, which construct the dual-frequency structure. For four-frequency case, the upper $F = 1, 2$ transition is driven by f2, f3 and f4 frequencies. For simplicity, there is a global detuning for three-frequency component model while for four components case we consider two kinds of detuning, which are named $\delta_1$, $\delta_2$, due to the specific level structure between the $F = 2$ and the upper $F = 1$ states. The f1, f2, f4 components share the same $\delta_1$ and the f3 has a separate $\delta_2$. A more general case with three or four different detunings instead of a global detuning would have little effect on the conclusions of the whole paper.

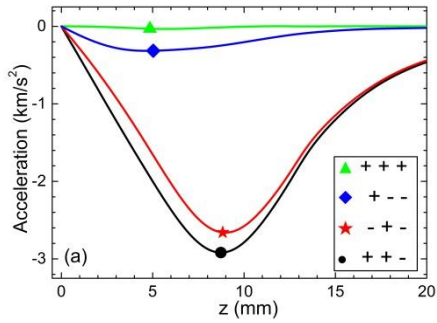
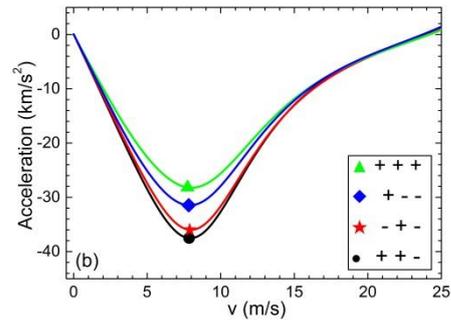
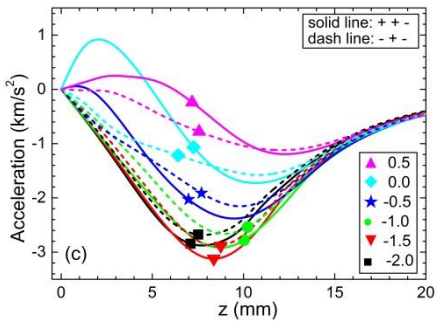
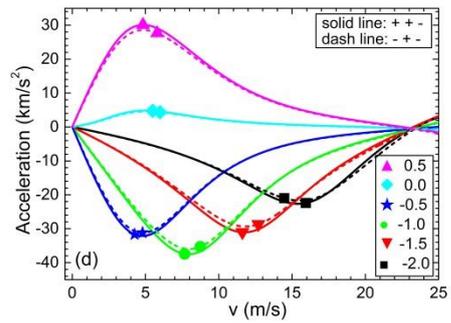

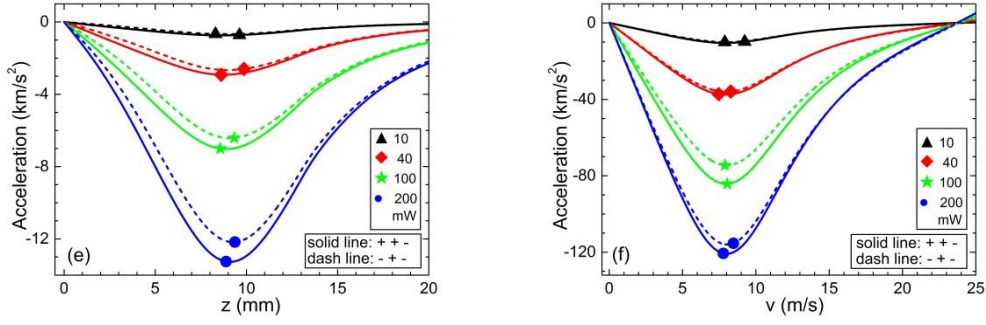

FIG. 4. Acceleration curves versus (a) displacement (b) speed, where the $A(0) - X(0)$ transition is driven using three frequency components. The values of the power for f1, f2 and f3 are 40, 40, and 80 mW, respectively. The detuning is $-\Gamma$. Acceleration versus (c) displacement (d) speed, for six different values of the detuning: $-2.0\Gamma$, $-1.5\Gamma$, $-1.0\Gamma$, $-0.5\Gamma$, $0.0\Gamma$ and $0.5\Gamma$. The solid line represents the (+ + -) configuration while the dash line is the (- + -) case. The power of each frequency component is the same as (a). Acceleration versus (e) displacement (f) speed, for four various values of the power for f1 frequency component: 10, 40, 100, 200 mW, where the power ratio of f1, f2 and f3 is 1 : 1 : 2. The detuning is $-\Gamma$.

## A. Three-frequency component model

Fig. 4 shows the results for the three-frequency case. Both the acceleration of a stationary molecule for a range of positions along the *z*-axis of the MOT and the accelerations for a range of velocities for a molecule at the origin, are obtained for various values of detuning and polarizations. If all sets of polarizations are reversed, the position dependent acceleration changes sign. There are four different configurations for three kinds of laser polarization. We denote the polarization of the frequency components as ($\pm$ $\pm$ $\pm$), for the case where f1, f2, f3 components have polarization $\sigma^\pm \sigma^\pm \sigma^\pm$, respectively.

From Fig.4 (a) and (b), we can see that the (+ + +) and (+ - -) configurations perform worse than the other two cases in restoring and damping force. This can be explained by the same polarization of f2 and f3, since they violate the dual-frequency mechanism. A molecule in dark states, relative to f3, is also not able to be pumped to cycling by f2. The minor differences between (- + -) and (+ + -) show that the main part of the force is supplied by $F = 2$ state, that is to say, the dual-frequency effect is the main mechanism responsible for the trapping force when the $A^2\Pi_{1/2} - X^2\Sigma^+$ transition is concerned. Note that though the f2 and f3 components can form restoring force for $F = 2$ level, they also generate the anti-restoring force for the upper $F = 1$ level. This is the reason that the total trapping force is less than half of that of the four-frequency configuration. In short, for three- frequency configuration, both (+ + -) and (- + -) appear good choices.

Fig 4(c) and (d) show how the acceleration depends on position and speed for various values of detuning. The solid line represents the (+ + -) configuration while the dash line is the (- + -) case. For (+ + -) scheme, when the detuning is positive, it heats the molecule and pushes the molecule away from the center. While the detuning is smaller than $-0.5\Gamma$, there is a net trapping and damping force and the optimum detuning is $-\Gamma$ after considering the two kinds of force comprehensively. On the other hand, the trapping acceleration of the (- + -) case is always exist throughout the values of detuning we investigated and the velocity dependent acceleration of the (- + -) scheme is almost overlapped with the (+ + -) one. The optimal detuning for (- + -) is also $-\Gamma$. To find the capture velocity of the MOT, $v_c$, we consider molecules enter the MOT in the *x-y* plane and are at 45° to the laser beams. We calculate the fastest speed a molecule can have if it is to be captured. Though the damping forces vary greatly with the global detuning, they almost pass through the same point with acceleration equal to zero, $v_c = 23$ m/s.

We also investigated the dependence of MOT force of MgF on the power of each frequency component. Since the saturation intensity of the MgF molecule is 62.5 mW/cm$^2$, which is much bigger than that of CaF and SrF, it can withstand greater laser intensity without oversaturation. The saturation parameters at 10, 40, 100, 200 mW are 0.07, 0.28, 0.7, 1.4, respectively. We only consider the values of laser power up to 200 mW, because the maximum output power of our laser system is ~ 1 W, which will be distributed to the four hyperfine levels. We can see from Fig. 4(e) and (f) that both the trapping and damping acceleration continue to increase as the laser power is increased up to 200 mW. We see that the damping force peaks when the speed is near 8m/s, corresponding to a Doppler shift that equals to the detuning of $-\Gamma$.

**B. Four-frequency component model**

The four-frequency case is complicated because there are more combinations and we set two kinds of detuning. To find the optimum polarization and detuning, we calculated the trapping frequency and the damping coefficient versus its detuning $\delta_1$ and $\delta_2$, for eight different polarization configurations. For simplicity, we only show the results of (+ + + -), which are in Fig. 5 (a) and (b). When $-1.0\Gamma < \delta_1 < 0$ and $-1.5\Gamma < \delta_2 < 0$, it provides a larger cooling force and if the detuning is positive, the result is almost opposite. For trapping molecules, $-0.75\Gamma < \delta_1 < 0.25\Gamma$ and $-0.25\Gamma < \delta_2 < 1.0\Gamma$ are the optimal ranges. By carefully seeking the overlapping areas, we obtain the optimum detuning: $\delta_1 = -0.5\Gamma$, $\delta_2 = -0.15\Gamma$. According to the results of our modeling, four kinds of polarization choices can provide relatively large force, which are (+ + + -), (- - + -), (+ - + -) and (- + + -), respectively. Then, we calculated the acceleration curves for these four configurations with the optimal detuning, as seen in Fig. 5 (c) and (d). In general, the damping force is one third smaller than that of the three-frequency case while the trapping force is about one and a half bigger than that of the three-frequency configuration. The (+ + + -) and (- + + -) perform better than that of (+ - + -) and (- - + -), since the trapping acceleration curves of the black rhombus and the green circle are lower than those of the blue pentagon and the red triangle for large displacement, which result in stronger acceleration. The insets in Fig. 5(c) and (d) are just for clarity. Also, the difference between (+ + + -) and (- + + -) configurations are too small to be measured by the experiment. However, for the convenience of experiment, the (+ + + -) configuration is preferred since the same $\sigma^+$ polarization could be addressed by one EOM at the same time. Similarly, we found the capture velocity is ~26 m/s.

Fig. 5(e) shows how the acceleration of a molecule depends on the displacement, for the laser power up to 200 mW. We can see that the peaks of the acceleration curve are pushed out from the center of the MOT with the increase of laser power. Fig. 5(f) shows how the acceleration depends on speed for various values of power, with detuning fixed to $\delta_1 = -0.5\Gamma, \delta_2 = -0.15\Gamma$. We find that the damping force peaks when the speed is near 5 m/s, which is a little bigger than the Doppler shift that equals to the detuning of $-0.5\Gamma$.

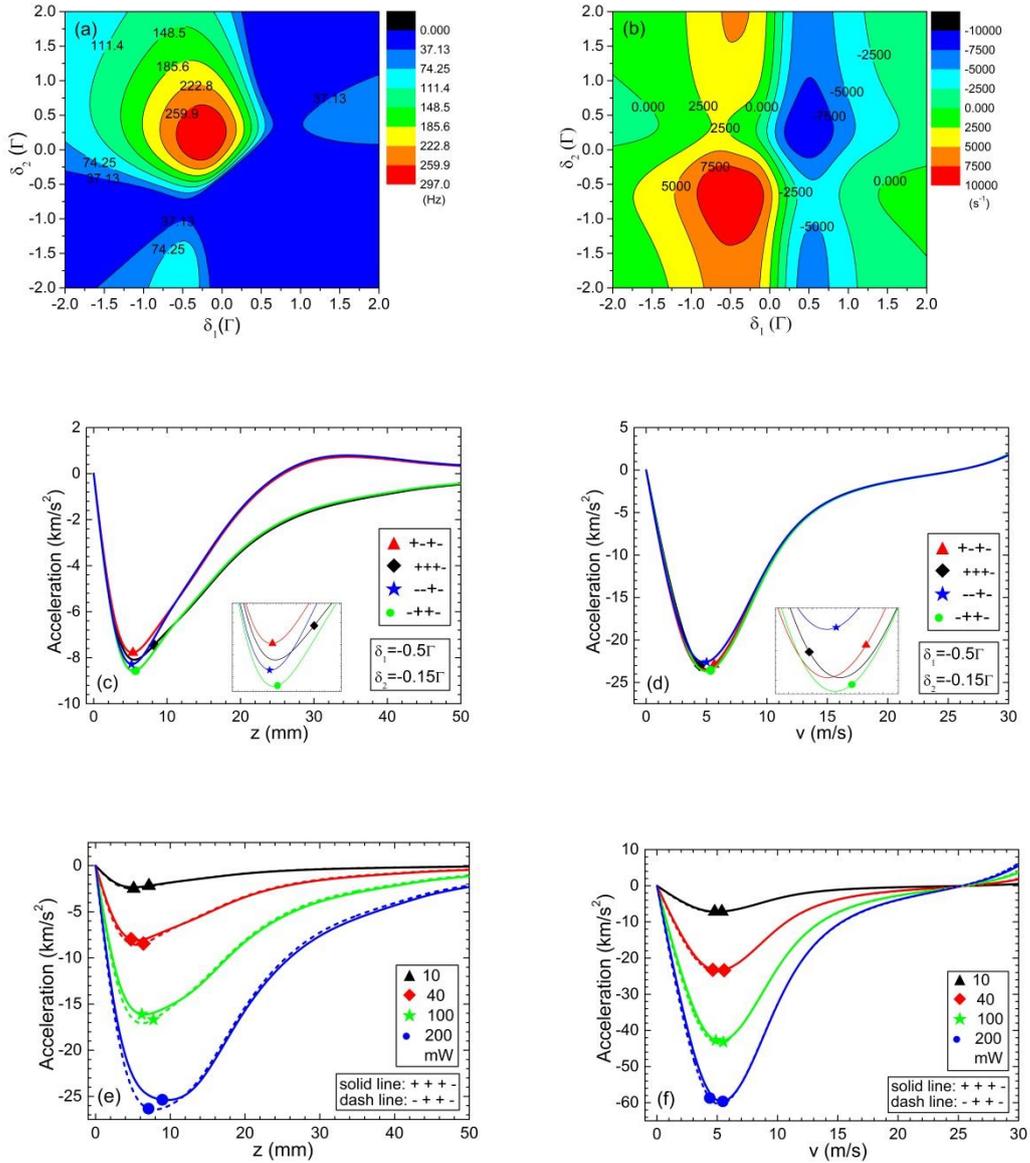

FIG. 5. (a) Trap frequency versus $\delta_1, \delta_2$ (b) Damping coefficient versus $\delta_1, \delta_2$ for (+ + + -) configuration. The optimal values of detuning are $\delta_1 = -0.5\Gamma, \delta_2 = -0.15\Gamma$. P = 40mW. Acceleration versus (c) displacement (d) speed, for four different polarization configurations: (+ - + -), (+ + + -), (- - + -) and (- + + -). The detuning: $\delta_1 = -0.5\Gamma, \delta_2 = -0.15\Gamma$ and the power of each frequency component is 40mW. Acceleration versus (e) displacement (f) speed, for four various values of the power in each MOT bean and each frequency component: 10, 40, 100, 200mW. The solid line represents the (+ + + -) configuration while the dash line is the (- + + -) case. The detuning: $\delta_1 = -0.5\Gamma, \delta_2 = -0.15\Gamma$.

**C. More-frequency component model**

Based on the above results, the trapping force of MgF molecule mainly depends on the dual-frequency arrangement of the $F = 2$ state. By carefully selecting laser polarization and detuning, both the trapping and damping force of MgF MOT are considerable despite of the tiny g-factor of the $A$ state. It is worthwhile considering whether the force can be further increased by applying the dual-frequency method to several of the hyperfine components. From Fig. 2, we can see that when $\delta_1 = -\Gamma$ and $\delta_2 = 2\Gamma$, both the trapping and damping force are strong. So, we can aim to arrange this situation for the other hyperfine components. Because the $F = 2$ and the upper $F = 1$ components are spaced by $0.4\Gamma$, we cannot have this situation for both of them at the same time. The $F = 0$ state has no Zeeman splitting and no dark state, so there is nothing to be gained from applying the two oppositely polarized frequency components. What remains is the lower $F = 1$ state. Because it has a negative g-factor ($g_g = -0.21$), as shown in Fig. 1, we can add one more frequency to address the lower $F = 1$ level for the (+ + -) configuration, which is detuned by $2\Gamma$ from this level and polarized $\sigma^-$. As for the (- + -) scheme, the same detuning and the opposite polarization would work. The specific set of three plus one frequencies and polarizations are given in the inset of Fig. 6(a), including four Zeeman sublevels of the ground state labeled with long solid line, the frequency components of (+ + -) configuration labeled with short solid line and the frequency components of (- + -) case with short dash line. For four-frequency component model, a frequency component with polarization $\sigma^-$ and detuning $2\Gamma$ is used to address the lower $F = 1$ state for the (+ + + -) configuration and a frequency component with the same detuning and the opposite polarization is arranged for the (- + + -) configuration, giving us the set of four plus one frequencies and polarizations illustrated in the inset of Fig. 6(c). Fig. 6(a) gives the acceleration versus position for the (+ + -) and (- + -) configurations. The addition of the extra component more than doubles the maximum acceleration to 7000 m/s$^2$ for (+ + -) case, while the acceleration for the (- + -) scheme is only increased to 4500 m/s$^2$. This is because the dual-frequency arrangement of the lower $F = 1$ state - constructed by the extra component - provides an anti-restoring force for the (- + -) configuration, though the whole force increases due to the increase of photon scattering. Fig. 6(b) shows the acceleration versus speed for the two cases. We find that the additional component has little influence on the maximum damping force or the damping coefficient. But, it narrows down slightly the range of velocities where the molecule can be cooled. The capture velocity decreases to $v_c = 17$ m/s. Fig. 6 (c) illustrated the acceleration versus position for the (+ + + -) and (- + + -) configurations. The variations of the trapping force for both (+ + + -) and (- + + -) configurations are small. Besides, neither the maximum damping force nor the damping coefficients change much, shown in Fig. 6(d). The capture velocity decreases to $v_c = 13$ m/s. These results suggest that a three plus one frequency component model, (- + + -), is able to provide a relatively large trapping force and the maximum damping force, which means the lowest temperature and the more molecules could be obtained for MgF.

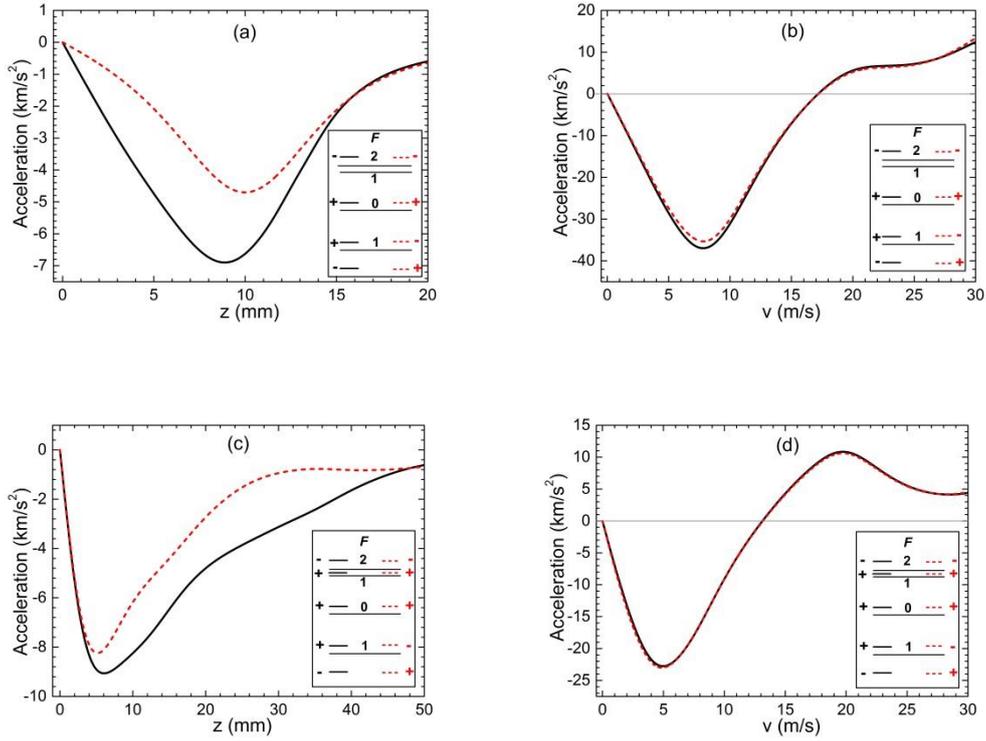

Fig. 6 Acceleration versus (a) displacement and (b) speed, for a MgF MOT operating on the $A(0)$ - $X(0)$ transition using three plus one frequency components. The values of the power for f1, f2 and f3 are 40, 40, and 80 mW, respectively. The power of the extra component is 40 mW. The detuning is $-\Gamma$ apart from the additional component whose detuning is $2\Gamma$. Acceleration versus (c) displacement and (d) speed, for a MgF MOT operating on the $A(0)$ - $X(0)$ transition using four plus one frequency components. The detuning: $\delta_1 = -0.5\Gamma, \delta_2 = -0.15\Gamma$, and the extra component is detuned by $2\Gamma$. The power of each frequency component is 40mW.

## IV. Conclusion

We have theoretically modeled the MOT force of MgF with 3D rate equations concerning the dual-frequency mechanism. We discuss some possible options for the polarizations and frequencies of the MOT, and suggest the optimized laser schemes for three- , four- and more-frequency configurations. Based on the modeling results, a three plus one frequency component scheme is suggested. In short, the MgF molecule is proved to be a good candidate for MOT despite of its tiny g-factor at the $A^2\Pi_{1/2}$ state.

**Acknowledgements**

We acknowledge the financial support from the National Natural Science Foundation of China under grants 11834003, 91836103, 91536218, 11374100, Natural Science Foundation of Shanghai Municipality under grant 17ZR1443000, Joint Research Institute for Science and Society (JoRISS), and the 111 project of China under Grant No. B12024.